\documentstyle[aps,epsfig,a4]{revtex}

\newcommand{\nc}{\nocite}
\newcommand{\beq}{\begin{equation}}
\newcommand{\eeq}{\end{equation}}
\newcommand{\bea}{\begin{eqnarray}}
\newcommand{\eea}{\end{eqnarray}}

\newcommand{\val}{{\rm val}}
\draft
\begin{document}
\title{Symmetry conserving approach to the quantization of the non-topological 
SU(3) soliton}
\author{M. Prasza{\l}owicz,\footnote{Alexander von Humboldt
Fellow, on leave from the Jagellonian University, Krak{\'o}w, Poland} 
T. Watabe\footnote{e-mail: watabe@hadron.tp2.ruhr-uni-bochum.de}
and K. Goeke}
\address{Institute for Theoretical Physics II, \\
Ruhr-University Bochum, \\
D-44780 Bochum, Germany }
\maketitle

\begin{abstract}
We reconsider canonical quantization of the rotating soliton in the SU(3)
Chiral Quark-Soliton Model. We show that at the level of $1/N_{\rm c}$, 
in contrast to the SU(2) version of the model, there appear terms which
spoil the commutation rules of the flavor generators. Terms of similar 
origin are also present in the expressions for axial couplings and magnetic
moments. We investigate the
small soliton limit of the model, and require that the results for the physical 
observables reduce to the ones of the non-relativistic quark model. This 
procedure allows to identify the troublesome terms. Next, we introduce a 
symmetry conserving approach by subtracting these terms and discuss numerical 
predictions obtained that way.
\end{abstract}

\vspace{1cm}



\vspace{-10cm}
\begin{flushright}
RUB-TPII-9/98 \\
hep-ph/9806431
\end{flushright}

\newpage

\section{Introduction}
\label{intro}

In this paper we are going to discuss the paradox, concerning the collective
quantization in the SU(3) Chiral Quark-Soliton Model ($\chi $QSM) \cite
{TWcharge}. In the SU(3) $\chi $QSM, in analogy with the Skyrme model
\cite{ANW}\nocite{JainWadia,Guad} -- \cite{MaNoPra},
one constructs the collective wave functions as the eigenstates of the 
symmetry generators such as flavor ${\cal T}_\alpha $ 
(or isospin ${\cal T}_A$ in the SU(2) version of the model)\footnote{%
Throughout this paper we assume that SU(3) indices, denoted by Greek
characters $\alpha$, $\beta \ldots$ run from 1$\ldots$8. The SU(2) subset of
these indices denoted by capital letters $A$, $B \ldots$ run from 1 to 3,
whereas indices denoted by $a$, $b \ldots$ run from 4 to 7.}, and spin
operators ${\cal S}_A$ \cite{SU3NJL}. 
There exists, however, the whole set of {\em %
generalized} spin generators including ${\cal S}_{4\ldots 7}$ and ${\cal S}_8
$ which form the SU(3) algebra. Precisely speaking $S_8$ is a constant equal
to $-N_{{\rm c}}/2\sqrt{3}$. It is however understood, that it is a constant
only on a subset of the entire Hilbert space, providing a constraint for the
physical states. This phenomenon is well known already from the SU(3) Skyrme
model, and it has been believed that there was no difference between the two
models as far as the the quantization procedure was concerned. In this paper
we will show that this equivalence actually breaks down.

All soliton models start from a classical, statical field configuration $U$
which is subsequently rotated by an SU(2) or SU(3) time-dependent matrix $%
A(t)$. Leaving the details for the next sections, let us quote one of the
most celebrated results of the SU(3) Skyrme model, namely the relation
between the {\em generalized} spin operators ${\cal S}$ (called often {\em %
right} generators) and the flavor ones ${\cal T}$ ({\em left} generators): 
\begin{equation}
{\cal T}_\alpha =-D_{\alpha \beta }(A)\;{\cal S}_\beta ,  \label{EqTS}
\end{equation}
where the summation over $\beta $ is assumed. Indeed ${\cal T}_\alpha $ have
proper commutation rules, provided ${\cal S}_\beta $ satisfy the SU(3)
algebra and the coefficients $D_{\alpha \beta }$ are Wigner matrices in the
adjoint representation of SU(3) (or SU(2)): 
\begin{equation}
A\lambda _\alpha A^{\dagger }=D_{\alpha \beta }(A)\;\lambda _\beta \;.
\label{Dalbet}
\end{equation}
Eq.(\ref{EqTS}) has been derived in the Skyrme model and it was believed to
be also true in the $\chi $QSM. In this paper we show that this relation is
violated in the SU(3) $\chi $QSM: 
\begin{equation}
{\cal T}_\alpha =-\left( 1+\frac 6{N_{{\rm c}}}\frac{I_2^{(-)}}{I_2^{(+)}}%
\right) D_{\alpha 8}(A)\;{\cal S}_8-\sum\limits_{\beta =1}^7D_{\alpha \beta
}(A)\;{\cal S}_\beta \;,  \label{TSviol}
\end{equation}
where $I_2^{(\pm )}$ are dynamical quantities defined in section \ref{toy}.
This is precisely the paradox: one constructs wave functions as
representations of flavor and {\em generalized} spin, however, there exists
a relation between the two sets of generators, which prevents one of them
(in our case flavor) from satisfying the proper commutation rules.

It is relatively easy to argue that the troublesome term $\sim I_2^{(-)}$
should be discarded, since our calculation was not complete as far as all $%
1/N_{{\rm c}}$ corrections are concerned. Indeed, one can imagine a number
of other contributions of this order which we have neglected in deriving Eq.(%
\ref{TSviol}). There is, however, one catch in this reasoning. Namely $S_8$
itself is of the order of $N_{{\rm c}}$, whereas all other $S_\beta $ in Eq.(%
\ref{TSviol}) are of the order of 1.  In fact SU(3) quantization also in the 
Skyrme model mixes
different orders in $1/N_{{\rm c}}$ expansion. Equation (\ref{EqTS}) itself
contains terms of different order in $N_{{\rm c}}$, nevertheless without the
subleading terms, namely ${\cal S}_{1\ldots 7}$ , the commutation rules for
flavor generators would not be satisfied either. For  more complicated
operators, as the charge operator for example, it is not possible to write
an analogue of Eq.(\ref{TSviol}) and clearly decide which terms should be
interpreted as non-leading $1/N_{{\rm c}}$ corrections and therefore
discarded, and which, although subleading, should be nevertheless retained.

There are two technical reasons  for the breakdown of the relation 
(\ref{EqTS}). The first one is that the SU(3) soliton 
models are not fully SU(3)
symmetric. Indeed, the SU(3) soliton is usually obtained by the, so called,
isospin embedding of the SU(2) soliton. This means that one of the SU(3)
generators, in our case $\lambda _8$, has a non-zero expectation value
between the soliton states. Let us note in passing, that because there is no
such operator in the two flavor case, the SU(2) version of the model does
not suffer from the quantization paradox \cite{WaWa,allstars}. 
The second reason consists in the
fact that $\chi $QSM has internal (quark) degrees of freedom which, in
contrast to the Skyrme model, give rise to the 
time-nonlocalities~\cite{allstars,BOrev} while calculating expectation values.
We will make all these statements clear in
section \ref{toy}, where a toy model formulation of the problem will be
presented. Earlier, in section \ref{guide}, we will briefly recapitulate
semiclassical quantization procedure and give literature where the details
can be found. In section \ref{path} we will reformulate the problem in terms
of the path integral.

Unfortunately, despite many efforts, theoretically satisfactory solution of
the quantization paradox has not yet been found. Therefore we would like to
propose a more modest approach. We will investigate a small soliton limit 
\cite{smallsol} of
the physical quantities of interest and require that they have limiting
values corresponding to the nonrelativistic quark model. This will allow us
to identify the troublesome $1/N_{{\rm c}}$ terms and analyze their
structure in terms of the quark degrees of freedom. Detailed analysis of the
charge operator, axial couplings and magnetic moments will be carried in
section \ref{smallsol}. For the purpose of the comparison with the data we
will subsequently discard these troublesome $1/N_{{\rm c}}$ terms, while 
retaining the other subleading contributions. (Symmetry conserving approach)
The results will be given in section \ref{numres}, where the model predictions 
{\em truncated} in this way will be
compared with experiment and also with the SU(2) version of the model, which
is free from the quantization ambiguities.

Although our procedure is in a sense arbitrary, it provides an insight into
the origin of the problem. Further support comes from the side of
phenomenology, since the truncated observables agree better (except for the
magnetic moments) with the data than the full ones. Moreover, for the
quantities which can be calculated also in the two flavor case, there is a
striking agreement between SU(2) and SU(3) results. We will summarize our
findings and present conclusions in section \ref{disc}.


\section{A short guide to the Chiral Quark Soliton Model}

\label{guide}

The generating functional of QCD involves quark and gluon fields. One can
imagine the following scenario
\footnote{See recent review by D.I. Diakonov \cite{DiakRev}}:
first integrate out gluons. The resulting action would then describe
the non-linear and non-local many quark interactions.
The next step would consists in linearizing this complicated action and 
expressing it in terms of local, color singlet composite fields
corresponding to the pseudo-scalar
mesons coupled in a chirally invariant way to the quark fields. And finally
we would have to integrate out quarks to end up with a pion 
(or $\pi $--K--$\eta $)
effective Lagrangian. So we go through a chain of effective actions: 
\begin{equation}
S_{{\rm QCD}}[q,\,A]\rightarrow S_{{\rm eff}}[q]\rightarrow S_{{\rm eff}%
}[q,\,\pi ,K,\eta ]\rightarrow S_{{\rm eff}}[\pi ,K,\eta ].  \label{eq:chain}
\end{equation}

It should be kept in mind that the arrows in Eq.(\ref{eq:chain}) do not
indicate a rigorous derivation of one action from another but rather {\em %
educated} guesses based mainly on symmetry principles and some physical
input. As $S_{{\rm eff}}[q,\pi ,K,\eta ]$ we will choose a semibosonized
action of the non-linear Nambu--Jona-Lasinio (NJL) model, which is follows
from QCD in the instanton liquid model of the QCD vacuum  \cite{dp}.
It is based on the following quark-meson Lagrangian (Chiral Quark-Soliton
Model): 
\begin{equation}
L=\bar{\psi}(i\gamma ^\mu \partial _\mu -\hat{m}-MU^{\gamma ^5})\psi \;,
\label{eq-lag}
\end{equation}
where $U^{\gamma _5}$ is defined by $U^{\gamma ^5}=\frac{1+\gamma _5}2U+%
\frac{1-\gamma _5}2U^{\dagger }$ and $M$ is the constituent quark mass of
the order 350--450~MeV. In what follows we will neglect current quark masses 
$\hat{m}$.

Next we will assume that baryons can be described as {\em solitons} 
\cite{dpp}\nc{RW} -- \cite{WY}
of the effective action corresponding to the Lagrangian (\ref{eq-lag}). That 
means that the Goldstone fields described by matrix $U$ are, in a sense, large 
and fulfill classical equations of motion. To this end we choose a {\em hedgehog }
Ansatz for $U$, which in the case of 2 flavors takes the following form
(see reviews: \cite{BOrev,TUErev}): 
\begin{equation}
U_2(\vec{r})=\exp \left( i\vec{\tau}\cdot \vec{n}\;F(r)\right) .  \label{U2}
\end{equation}
Here $F(r)$ is a profile function, and $\vec{n}=\vec{r}/r$ is a unit vector.
In SU(3) case the soliton field reads: 
\begin{equation}
{U_3}(\vec{r})=\left( 
\begin{array}{cc}
U_2 & 0 \\ 
0 & 1
\end{array}
\right) \ .
\label{U3}
\end{equation}
Obviously  ${U_3}$ is not completely symmetric in the SU(3) flavor space;
namely it commutes with $\lambda _8$.

Next one allows for the rotation in the flavor space: $U(\vec{r})\rightarrow
A(t)U(\vec{r})A^{\dagger }(t)$ with angular velocity\footnote{%
Obviously in the SU(2) case index $\alpha$ is confined to 1$\ldots$3 and
Gell-Mann matrices $\lambda$ should be replaced by Pauli matrices $\tau$.}: 
\begin{equation}
\Omega _\alpha =-i{\rm Tr}\left( \lambda _\alpha \,A^{\dagger }\frac d{dt}%
A\right) .
\end{equation}
The Lagrangian of the system takes then the following form: 
\begin{equation}
L=\psi ^{\dagger }A\left( i\partial _t+i\vec{\alpha}\cdot \vec{\nabla}-\beta
M\;{U}^{\gamma ^5}-\frac 12\lambda _\alpha \Omega _\alpha \right) A^{\dagger
}\psi =\psi ^{\dagger }A\;D[U,\Omega ]\;A^{\dagger }\psi \ .  \label{Ddef}
\end{equation}

The detailed path integral derivation of the collective Hamiltonian
describing baryon spectrum will be given in section \ref{path}, where the
subtleties connected with the non-local nature of the effective action will
be discussed. By integrating out the quark fields one arrives at the {\em %
collective} Lagrangian ${\cal L}$, which is a power series in $\Omega $: 
\begin{equation}
{\cal L}=-\frac{N_{{\rm c}}}{2\sqrt{3}}\Omega _8+\frac 12I_1^{AB}\;\Omega
_A\Omega _B+\frac 12I_2^{ab}\;\Omega _a\Omega _b+\ldots \;,  \label{collL}
\end{equation}
where matrices $I_{1,2}$ are the inertia tensors in the subspaces $1\ldots 3$
and $4\ldots 7$ respectively. Normal procedure consists in constructing
effective Hamiltonian ${\cal H}$ acting in the representation space spanned
by wave functions, which transform as irreducible representations of two
symmetry groups: 1) left multiplication of $A$ by an SU(3) matrix and 2)
right multiplication of $A$ by an SU(2) matrix. The left symmetry
corresponds to flavor, whereas the right one to spin. In fact one can
promote the right symmetry group to the full SU(3) and then consider only
the subset of the full Hilbert space, for which the right hypercharge is
equal to $-1$. This constraint follows from the fact that $\Omega _8$
appears only linearly in the collective Lagrangian (\ref{collL}). Therefore
the baryon wave function of flavor\footnote{%
Here three numbers in brackets correspond to the SU(3) states of the {\em %
left} ($B$) or {\em right} ($S$) symmetry group.} $B=(YTT_3)$ and spin $%
S=(-1SS_3)$ can be explicitly written in terms of the SU(3) Wigner $D^{(%
{\cal R})}$ functions: 
\begin{equation}
\Psi _{BS}^{({\cal R)}}\;=\;(-)^{S_3-1/2}\sqrt{\mbox{dim}({\cal R)}}\left[
D_{(YTT_3)(1S-S_3)}^{({\cal R)}}\right] ^{*}.  \label{wf}
\end{equation}
Note that the lowest SU(3) representations ${\cal R}$ allowed by the
constraint $Y_{{\rm R}}=-N_{{\rm c}}/3$ correspond to octet and decuplet for
$N_{{\rm c}}=3$.

In section \ref{path} we will show in more detail how to calculate tensors
of inertia and expectation values of various operators in the path integral
formalism. Before doing that, let us consider a simple quantum-mechanical
model, which reveals all necessary features needed to state the quantization
paradox.


\section{Quantum-mechanical toy model}

\label{toy}

Consider\footnote{We thank W. Broniowski for pointing to us this simple
way of stating the problem.} 
a relativistic Hamiltonian $H[U]$, corresponding to the Lagrangian (%
\ref{eq-lag}), satisfying a single-particle eigen-equation:
\begin{equation}
H[U] \; |\; i\; \rangle = \varepsilon_i |\; i\; \rangle \;.  \label{spectrum}
\end{equation}
In Eq.(\ref{spectrum}) the color index of the quark fields has been
suppressed. As usually we assume that the ground state consists of all
Dirac see levels and one positive energy level, called {\em valence}, filled.

Rotation induces a perturbation into the Hamiltonian (\ref{spectrum}): 
\begin{equation}
H[U] \rightarrow A \left( H[U] - \frac{1}{2} \lambda_{\alpha}
\Omega_{\alpha} \right) A^{\dagger} = A H^\prime [U]A^{\dagger} .
\end{equation}
For further calculations it is convenient to work in the rotating frame: 
\begin{equation}
|\; \tilde{i} \; \rangle = A^{\dagger} \; |\; i \; \rangle,
\end{equation}
where the eigen-equation reads: 
\begin{equation}
\left( H[U] - \frac{1}{2} \lambda_{\alpha} \Omega_{\alpha} \right) | \; 
\tilde{i} \; \rangle = \tilde{\varepsilon}_i |\; \tilde{i}\; \rangle.
\label{Hprime}
\end{equation}
We will solve Eq.(\ref{Hprime}) perturbatively, noting that the unperturbed
spectrum satisfies Eq.(\ref{spectrum}). In the first order we get: 
\begin{equation}
| \;\tilde{i} \; \rangle = \left( | \; i \; \rangle - \frac{1}{2}
\sum\limits_{j \ne i} | \; j\; \rangle \frac{\langle \;j\;| \lambda_\alpha
\Omega_\alpha | \; i \; \rangle} {\varepsilon_j - \varepsilon_i} \right) \;.
\label{perturbed}
\end{equation}

Now in order to calculate the matrix element of any {\em intrinsic} operator
consisting of a string of $\gamma $ matrices (and possibly some space
dependent function) $\Gamma $ and a flavor matrix $\lambda _\alpha $: 
\begin{equation}
O_\alpha ^\Gamma =\lambda _\alpha \Gamma   \label{O}
\end{equation}
we have first to transform it into the rotating frame: 
\begin{equation}
O_\alpha ^\Gamma \rightarrow \tilde{O}_\alpha ^\Gamma =AO_\alpha ^\Gamma
A^{\dagger }=D_{\alpha \rho }\lambda _\rho \;\Gamma \;.  \label{Orotated}
\end{equation}
Operator $\tilde{O}_\alpha $ has to be subsequently sandwiched between the
occupied states (\ref{perturbed}) and summed over $i$. In this way we get
expression for the collective operator ${\cal O}_\alpha $ which acts in the
Hilbert space of the collective baryon states given in Eq.(\ref{wf}): 
\begin{eqnarray}
{\cal O}_\alpha ^\Gamma  &=&\sum_{\scriptstyle m\in {\rm occ.}}\langle
m|D_{\alpha \beta }\lambda _\beta \Gamma |m\rangle   \nonumber \\
&-&\frac 12\sum_{
\begin{array}{c}
\scriptstyle m\in {\rm occ.} \\ 
\scriptstyle n\in {\rm non-occ.}
\end{array}
}\frac{\langle m|D_{\alpha \rho }\lambda _\rho \Gamma |n\rangle \langle
n|\lambda _\beta \Omega _\beta |m\rangle +\langle m|\lambda _\beta \Omega
_\beta |n\rangle \langle n|D_{\alpha \rho }\lambda _\rho \Gamma |m\rangle }{%
\varepsilon _n-\varepsilon _m}\;.  \label{Ocoll}
\end{eqnarray}
Throughout this paper we shall always assume that $n$ runs over non-occupied
states, whereas $m$ runs over occupied states (including the valence level).
Of course the above formula for ${\cal {O}_\alpha ^\Gamma }$ requires
regularization which does not concern us for the moment.

Note that in Eq.(\ref{Ocoll}) matrix $D_{\alpha \rho }(A)$ and angular
velocity $\Omega _\beta $ appear in two different orders. We are not going
to commute them because in the process of collective quantization $\Omega
_\beta $ will be promoted to the angular momentum operators, which do not
commute with $D_{\alpha \rho }(A)$. Although it may seem that there is some
arbitrariness in the way we define ${\cal {O}_\alpha ^\Gamma }$ as far as
the ordering of $D_{\alpha \rho }(A)$ and $\Omega _\beta $ is concerned, we
will stick to the natural order dictated by the perturbative expansion of
the rotated levels. We shall argue in the next section that in the path
integral formulation the order of the operators is actually unambiguous.

Now using decomposition into commutator and anticommutator: 
\begin{eqnarray}
D_{\alpha \rho} \Omega_\beta & = & ~~\frac{1}{2} \left[ D_{\alpha \rho},
\Omega_\beta \right] + \frac{1}{2} \left\{D_{\alpha \rho}, \Omega_\beta
\right\},  \nonumber \\
\Omega_\beta D_{\alpha \rho}& = & - \frac{1}{2} \left[ D_{\alpha \rho},
\Omega_\beta \right] + \frac{1}{2} \left\{D_{\alpha \rho}, \Omega_\beta
\right\}  \label{DOm}
\end{eqnarray}
we can rewrite Eq.(\ref{Ocoll}) as follows: 
\begin{eqnarray}
{\cal O}_\alpha^\Gamma &=& N_{{\rm c}} D_{\alpha \beta} \sum\limits_m
\langle m | \lambda_{\beta} \Gamma | m \rangle  \nonumber \\
&-& \frac{N_{{\rm c}}}{4} \left\{ D_{\alpha \rho}, \Omega_\beta \right\}
\sum\limits_{m,n} \frac{\langle m | \lambda_{\rho} \Gamma| n \rangle \langle
n | \lambda_\beta | m \rangle + \langle m | \lambda_\beta | n \rangle
\langle n | \lambda_{\rho} \Gamma | m \rangle } {\varepsilon_n -
\varepsilon_m}  \nonumber \\
&-& \frac{N_{{\rm c}}}{4} \left[ D_{\alpha \rho}, \Omega_\beta \right]
\sum\limits_{m,n} \frac{\langle m | \lambda_{\rho} \Gamma| n \rangle \langle
n | \lambda_\beta | m \rangle - \langle m | \lambda_\beta | n \rangle
\langle n | \lambda_{\rho} \Gamma | m \rangle } {\varepsilon_n -
\varepsilon_m} \;,  \label{Ocoll1}
\end{eqnarray}
where we have explicitly pulled out an overall factor $N_{{\rm c}}$
corresponding to the sum over colors.

Equation (\ref{Ocoll1}) is our final result for any observable. Let us first
consider collective isospin generators in the SU(2) version of the model
(with obvious replacement of Gell-Mann matrices $\lambda $ by Pauli matrices 
$\tau $). Of course we expect to reproduce Eq.(\ref{EqTS}) in this case. For
the isospin $\Gamma =1/2$ and the first term in Eq.(\ref{Ocoll1}) is zero
since the SU(2) hedgehog is fully symmetric. For the same reason the last
term vanishes and we are left with the simple formula: 
\begin{equation}
{\cal T}_A=-\frac 12I_1^{(+)}\;\left\{ D_{AB},\Omega _B\right\} \;
\label{TA}
\end{equation}
where the SU(2) moment of inertia is defined through the following relation: 
\begin{equation}
\frac{N_{{\rm c}}}4\sum\limits_{m,n}\frac{\langle m|\tau _C|n\rangle \langle
n|\tau _B|m\rangle +\langle m|\tau _B|n\rangle \langle n|\tau _C|m\rangle }{%
\varepsilon _n-\varepsilon _m}=\delta _{BC}\;I_1^{(+)}.  \label{I1plus}
\end{equation}

In order to reproduce Eq.(\ref{EqTS}) we have to define spin generators. We
shall do this first for SU(3) and then restrict the general formula to the
SU(2) case. To this end let us observe that the total angular momentum $J_A$
is defined {\em ab initio} in the rotating frame. Therefore there is no need
to transform spin generators\footnote{%
One should remember that $J_A$, which correspond to the total angular
momentum of the soliton, are subsequently quantized as spins of the
collective states. That is why we interchangeably call them spins or total
angular momentum operators, which, however, should not cause any confusion.}
to the rotating frame, as it was done in Eq.(\ref{Orotated}). Because of the 
{\em hedgehog} symmetry the matrix elements of $J_A$ are equal to the matrix
elements of $-1/2\;\lambda _A$, which allows us to define {\em generalized}
spin generators corresponding to all eight $\lambda $ matrices. In fact we
can simply use Eq.(\ref{Ocoll1}) with $\Gamma =1/2$, replacing $D_{\alpha
\rho }\rightarrow -\delta _{\alpha \rho }$, which gives: 
\begin{equation}
{\cal S}_\alpha =-\frac{N_{{\rm c}}}2\sum\limits_m\langle m|\lambda _\alpha
|m\rangle +\frac{N_{{\rm c}}}4\Omega _\beta \sum\limits_{m,n}\frac{\langle
m|\lambda _\alpha |n\rangle \langle n|\lambda _\beta |m\rangle +\langle
m|\lambda _\beta |n\rangle \langle n|\lambda _\alpha |m\rangle }{\varepsilon
_n-\varepsilon _m}\;.  \label{Salpha}
\end{equation}
Note that the antisymmetric part (the third term in Eq.(\ref{Ocoll1})) does
not contribute here.

Let us turn to the SU(2) case. Again the first term in Eq.(\ref{Salpha})
vanishes and in the second term we immediately recognize the moment of
inertia $I_1^{(+)}$ defined in Eq.(\ref{I1plus}). Finally, due to the
commutation rule 
\begin{equation}
\left[ {\cal S}_\beta, D_{\alpha \rho} \right] = i f_{\beta \rho \gamma}
D_{\alpha \gamma} \;\;\; {\rm or} \;\;\; \left[ {\cal S}_B, D_{A R} \right]
= i \epsilon_{B R C} D_{A C}  \label{SDcomm}
\end{equation}
the order of the operators in $D_{AB}$ and ${\cal S}_B$ (summed over $B$)
in the anticommutator in Eq.(\ref{TA}) does not matter and one arrives, 
as expected, at Eq.(\ref{EqTS}).

What goes wrong in the SU(3) case? Let us first observe that Eq.(\ref
{spectrum}) separates into two independent parts: one corresponding to the
subspace of the SU(2) soliton (see Eq.(\ref{U3})) and the plane wave sector
corresponding to the ``strange'' quark\footnote{%
This nomenclature is not entirely correct, since we work in the rotating
frame, where all quark flavors are mixed.}. Plane waves with negative energy
which are occupied will be denoted by $m_0$, whereas the non-occupied
positive energy levels will be denoted as $n_0$.

\begin{figure}[t]
\centerline{
\epsfig{file=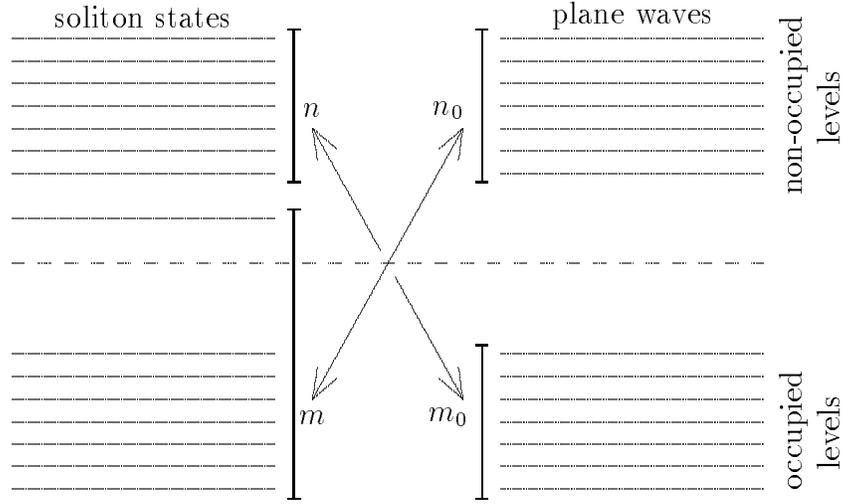}
}
\caption{Spectrum of the Dirac Hamiltonian (12).}
\end{figure}

Let us first calculate ${\cal S}_\alpha$. To this end let us observe that
the first term in Eq.(\ref{Salpha}) does not vanish since $\lambda_8$ has a
non-vanishing expectation value: 
\begin{equation}
\Lambda^\alpha \equiv \frac{N_{{\rm c}}}{2} \sum\limits_m \langle m |
\lambda_{\alpha} | m \rangle =\frac{N_{{\rm c}}}{2 \sqrt{3}}\;
\delta_{\alpha 8},  \label{Lamal}
\end{equation}
provided one subtracts contribution of the vacuum. Next, the sum multiplying 
$\Omega_\beta$ splits into two independent parts: 
\begin{equation}
\frac{N_{{\rm c}}}{4} \sum\limits_{m,n} \frac{\langle m | \lambda_{\alpha}|
n \rangle \langle n | \lambda_\beta | m \rangle + \langle m | \lambda_\beta
| n \rangle \langle n | \lambda_{\alpha}| m \rangle } {\varepsilon_n -
\varepsilon_m} = \delta_{AB} \; I_{1}^{(+)} + \delta_{ab}\; I_2^{(+)} \; .
\label{I2plus}
\end{equation}
This equation defines $I_2^{(+)}$, which can be reexpressed in terms of the
overlaps between the soliton states and the plane wave states: 
\begin{equation}
I_2^{(+)}= \frac{N_{{\rm c}}}{4}\left\{ \sum\limits_{m,n_0} \frac{\langle m
| n_0 \rangle \langle n_0 | m \rangle}{\varepsilon_{n_0} - \varepsilon_m} +
\sum\limits_{m_0,n} \frac{\langle m_0 | n \rangle \langle n | m_0 \rangle } {%
\varepsilon_n - \varepsilon_{m_0}} \right\} \; .  \label{I2plus1}	
\end{equation}
These transitions ale illustrated in Fig.1.

Throughout this paper we use the subscript ''1'' in the definitions of the
inertia parameters, if only the soliton states are involved, while if
subscript ''2'' is used, it is understood that transitions between the plane
waves and the soliton states also contribute. Superscripts ''$(\pm )$'' are
used to denote symmetrized or antisymmetrized contributions. Clearly $%
I_1^{(-)}\equiv 0$, because the soliton is totally symmetric in the SU(2)
subspace.

Finally we get the expression for ${\cal S}_{\alpha}$: 
\begin{equation}
{\cal S}_{\alpha} = -\frac{N_{{\rm c}}}{2 \sqrt{3}} \delta_{\alpha 8}+
I_{1}^{(+)} \; \Omega_A + I_2^{(+)} \; \Omega_a \; .  \label{Salpha1}
\end{equation}

Now we are ready to calculate flavor generators. Substituting $\Gamma =1/2$
in Eq.(\ref{Ocoll}) we observe that: 
\begin{equation}
{\cal T}_\alpha =-D_{\alpha \beta }(A)\;{\cal S}_\beta -\frac{N_{{\rm c}}}8%
\left[ D_{\alpha \rho },\Omega _\beta \right] \sum\limits_{m,n}\frac{\langle
m|\lambda _\rho |n\rangle \langle n|\lambda _\beta |m\rangle -\langle
m|\lambda _\beta |n\rangle \langle n|\lambda _\rho |m\rangle }{\varepsilon
_n-\varepsilon _m}\;.  \label{TS1}
\end{equation}
This is precisely the last term in Eq.(\ref{TS1}) which causes the problem.
Were $\Omega $'s the classical quantities, the commutator would vanish.

It is easy to convince oneself that the last term in Eq.(\ref{TS1}) is unequal 
to zero only for the indices $\rho =r$ and $\beta =b$ , with  $r,b=4\ldots 7$: 
\begin{equation}
\frac{N_{{\rm c}}}4\sum\limits_{m,n}\frac{\langle m|\lambda _r|n\rangle
\langle n|\lambda _b|m\rangle -\langle m|\lambda _b|n\rangle \langle
n|\lambda _r|m\rangle }{\varepsilon _n-\varepsilon _m}=i\frac 2{\sqrt{3}}%
f_{8rb}\;I_2^{(-)}\;.  \label{I2minus}
\end{equation}
Again, $I_2^{(-)}$ can be rewritten in terms of the overlaps between the
plane waves and the soliton states: 
\begin{equation}
I_2^{(-)}=\frac{N_{{\rm c}}}4\left\{ \sum\limits_{m,n_0}\frac{\langle
m|n_0\rangle \langle n_0|m\rangle }{\varepsilon _{n_0}-\varepsilon _m}%
-\sum\limits_{m_0,n}\frac{\langle m_0|n\rangle \langle n|m_0\rangle }{%
\varepsilon _n-\varepsilon _{m_0}}\right\} \;.  \label{I2minus1}
\end{equation}
It is now clear that for the fully symmetric Ansatz $I_2^{(-)}$ would
vanish. Indeed, in that case the plane wave spectrum would have to be
replaced by the soliton spectrum, and the two terms in Eq.(\ref{I2minus1})
would cancel. This is precisely the reason why there is no antisymmetric
piece in the SU(2) case ({\em i.e.} for $\rho =R$ and $\beta =B$ , with $%
R,B=1\ldots 3$).

The final expression for the flavor generators is obtained by contracting  $f
$ \ symbols: 
\begin{equation}
f_{8bc}\,f_{bc\rho }=3\,\delta _{8\rho },  \label{ff1}
\end{equation}
yielding: 
\begin{equation}
{\cal T}_\alpha =-D_{\alpha \beta }(A)\;{\cal S}_\beta +\sqrt{3}D_{\alpha
8}(A)\frac{I_2^{(-)}}{I_2^{(+)}}\;.  \label{TS2}
\end{equation}

A few remarks concerning the last term in Eq.(\ref{TS2}) are in order.
First, this term spoils commutation rules for the flavor generators.
Moreover it is non-diagonal in the SU(3) representation space. Finally, it
would vanish for the symmetric spectrum, as clearly seen from Eq.(\ref
{I2minus1}) and Fig.1.

Using relation (\ref{TS2}) one can calculate the nucleon electric charge by
sandwiching the charge operator between the hadronic states of Eq.(\ref{wf}%
): 
\begin{eqnarray}
e_N = \left( T_3 + \frac{1}{2} Y \right) +\frac{1}{5}\left( T_3 + \frac{3}{2}
\right) \frac{I_{2(-)}}{I_{2(+)}} \ ,  \label{eq-ecbk}
\end{eqnarray}
where $T_3$ stands for the isospin of the nucleon.

Finally, let us note that generally for $\Gamma \ne 1$ there will be
antisymmetric contributions in Eq.(\ref{Ocoll1}) both within  the soliton
states ({\em i.e.} for $\rho =R,\;\beta =B$, with $R,B=1\ldots 3$) and for 
the transitions between the soliton states and the plane waves.


\section{Path integral formulation of the problem}

\label{path}

\subsection{Collective Lagrangian and adiabatic approximation}

In this section we will rederive Eq.(\ref{collL}) and Eq.(\ref{Ocoll1})
within the path integral formalism. As we shall see, the path integral will
uniquely determine the order of the $D_{\alpha \rho }$ functions and angular
velocities $\Omega _\beta $. To this end let us consider the 3rd action from
the chain of Eq.(\ref{eq:chain}), with the Lagrangian density given by Eq.(%
\ref{eq-lag}). The pertinent partition function reads: 
\begin{equation}
{\cal Z}=\int {\cal D}\psi ^{\dagger }{\cal D}\psi {\cal D}U\ e^{i\int d^4x\;%
{L}}\ .
\label{ourS}
\end{equation}

Let us first outline steps needed to derive the collective Lagrangian (\ref
{collL}). For this purpose one considers the nucleon correlation function: 
\begin{equation}
\Pi _N(\vec{x},T/2;\vec{y},-T/2)=\left\langle 0\left| J_N(\vec{x}%
,T/2)J_N^{\dagger }(\vec{y},-T/2)\right| 0\right\rangle \ ,
\label{corrdef}
\end{equation}
where the $J_N(\vec{x},t)$ and $J_N^{\dagger }(\vec{x},t)$ are nucleon
annihilation and creation operators, respectively. 
It is implicitly assumed that the nucleon consists of $N_{{\rm c}}$ rather
than of 3 quarks. Although we work in the Minkowski space-time we will take
the limit $iT\rightarrow \infty $ in order to select the ground state
contribution to the correlator (\ref{corrdef}): 
\begin{equation}
\lim\limits_{iT\rightarrow \infty }\Pi _N(\vec{x},T/2;\vec{y},-T/2)\sim \exp
(-iTE_{{\rm cl}})\ .
\label{corrlim}
\end{equation}
Within the path integral formalism Eq.(\ref{corrdef}) reads: 
\[
\Pi _N(\vec{x},T/2;\vec{y},-T/2)=\frac 1{{\cal Z}}\int {\cal D}\psi
^{\dagger }{\cal D}\psi {\cal D}U\ J_N(\vec{x},T/2)\,e^{i\int d^4x^{\prime }{%
L}}\ J_N^{\dagger }(\vec{y},-T/2)\ .
\]
The integration over the quark fields can be carried out exactly. In order
to integrate over the chiral meson field $U$ we first introduce a classical
solution, the {\it soliton}, which is given by a dominant trajectory in the
large $N_{{\rm c}}$ limit. The classical equation of motion for ${U}$ is
derived from the variation of the classical energy $E_{{\rm cl}}$ in Eq.(\ref
{corrlim}). In this way the soliton profile function $F$ and the soliton
mass ${\cal M}$ are calculated.

Next, we take only zero-frequency modes of $U$ around the classical
solution, {\it i.e.} the rotational modes\footnote{%
We neglect here translational mode which would restore translational
invariance.}. Then, the path integral ${\cal D}U$ is replaced by ${\cal D}%
\xi $, where $\xi _\alpha $ are Euler angles corresponding to the rotation
matrix $A(t)=A(\xi (t))$: 
\begin{equation}
\int d^3xd^3y\ \Pi _N(\vec{x},T/2;\vec{y},-T/2)=\int {\cal D}\xi \ \Psi
_N^{\dagger }(\xi (T/2))\ e^{N_{{\rm c}}{\rm Sp}\log D[{U},\xi ]}\ \Psi
_N(\xi (-T/2))\ ,
\label{corrD}
\end{equation}
where Sp stands for the functional and matrix trace of the logarithm of the
Dirac operator $D[U,\xi ]$ defined in Eq.(\ref{Ddef}). The $\Psi _N(\xi )$
is a nucleon collective wave function, which is a function of the Euler
angles $\xi _\alpha $. It is normalized by: 
\[
\int d\xi \ \Psi _{N^{\prime }}^{\dagger }(\xi )\Psi _N(\xi )=\delta
_{N^{\prime }N}\ .
\]
It is already clear from Eq.(\ref{corrD}) that the effective action (which,
by the way, corresponds to the last one in Eq.(\ref{eq:chain})) is highly
nonlocal. This fact is further visible if one expands ${\rm Sp}\log D$ in
powers of the angular velocity $\Omega $ up to $O(\Omega ^2)$. This is
pictorially illustrated in Fig.2 where the quark loop with the full
propagator is expanded into the powers of $\Omega $ represented by crosses.

\begin{figure}[h]
\centerline{
\epsfig{file=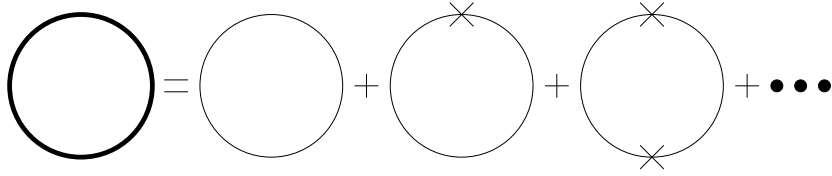,width=11cm}
}
\caption{Full quark loop (thick line) expanded in powers of angular velocity
$\Omega$ (denoted by crosses).}
\end{figure}

Because of the time dependence of $\Omega $, higher order terms give a
nonlocality in time: 
\begin{eqnarray}
\int d^3xd^3y\ \Pi _N(\vec{x},T/2;\vec{y},-T/2) &=&\int {\cal D}\xi \ \Psi
_N^{\dagger }(\xi (T/2))\ e^{-iE_{{\rm cl}}\;T+\int dt_1dt_2{\cal L}_{{\rm NL%
}}(t_1,t_2)}\ \Psi _N(\xi (-T/2))\ ,  \nonumber \\
~ &~&~
\end{eqnarray}
The collective nonlocal Lagrangian ${\cal L}_{{\rm NL}}(t_1,t_2)$ should be
understood as: 
\begin{eqnarray}
\int dt_1dt_2\ {\cal L}_{{\rm NL}}(t_1,t_2) &=&i\;\int dt_1dt_2\left\{
-\Lambda ^\alpha \Omega _\alpha (t_1)\ \delta (t_1-t_2)\right.   \nonumber \\
&&\left. +\frac 12\Omega _\alpha (t_1)\;I^{\alpha \beta }(t_1-t_2)\;\Omega
_\beta (t_2)\ \Theta (t_1-t_2)\right\} \ ,  \label{DcollL}
\end{eqnarray}
where 
\[
I^{\alpha \beta }(t)=i\frac{N_{{\rm c}}}2\sum\limits_{mn}e^{-i(\varepsilon
_n-\varepsilon _m)t}\langle m|\lambda _\alpha |n\rangle \langle n|\lambda
_\beta |m\rangle \ ,
\]
where states $n$ and $m$ have been already introduced in the previous
section and $\Lambda ^\alpha $ is given in Eq.(\ref{Lamal}).

In order to recover the local Lagrangian (\ref{collL}) we have to make an
adiabatic approximation and assume that $\Omega_\alpha$ very weakly depends
on time. This amounts to: $\Omega_\alpha(t_2) \approx \Omega_\alpha(t_1)$.
Hence\footnote{%
Local and nonlocal inertia parameters can be distinguished by the explicit
time dependence in the latter case}: 
\begin{eqnarray}
\int dt_1 dt_2 \ \Omega_\alpha(t_1) \; I^{\alpha \beta}(t_1-t_2) \;
\Omega_\alpha(t_2) \ \Theta(t_1-t_2) \approx ~~~~~~~~~~~~~~~  \nonumber \\
~~~~~~~~~~~~~~~ \int dt_1 \ \Omega_\alpha(t_1) \left\{ \int dt_2 I^{\alpha
\beta}(t_1-t_2) \Theta(t_1-t_2) \right\} \Omega_\alpha(t_1) \ .
\label{adiabatic}
\end{eqnarray}
The integration of the $I^{\alpha \beta}(t_1-t_2)$ over $t_2$ gives: 
\begin{eqnarray}
I^{\alpha \beta} = \lim \limits_{iT \rightarrow \infty} \int\limits_0^{\frac{%
T}{2}-t_1} dt \ I^{\alpha \beta} (t) &=& \frac{N_{{\rm c}}}{2} \sum_{m,n} 
\frac{\langle m|\lambda_\alpha |n \rangle \langle n|\lambda_\beta |m \rangle%
} {\varepsilon_n-\varepsilon_m} \; .  \label{Ialbe}
\end{eqnarray}
Note that this integration is finite only if $\varepsilon_m < \varepsilon_n$%
. This condition and the Pauli principle make that only transitions from
occupied $m$ to nonoccupied $n$ states are possible.

Needless to say that formula (\ref{Ialbe}) contains both symmetric and
antisymmetric contributions. Symmetric contribution has been defined in Eqs.(%
\ref{I1plus}, \ref{I2plus}), whereas the antisymmetric one is given by (\ref
{I2minus}). However, due to the symmetricity of the product $\Omega_\alpha
\Omega_\beta$ the antisymmetric part does not contribute to the collective
Lagrangian.

The nonlocality in time encountered here is relatively harmless. By
employing adiabatic approximation we recover the local collective Lagrangian
(\ref{collL}) and the collective quantization can be carried through in the
normal way. Let us also mention that in many cases one employs a different
procedure to calculate moments of inertia using the identity for the real
part ${\rm Re}~{\rm log}D=1/2\;{\rm log}D^{\dagger }D$. This method,
although useful {\em e.g.} in proper time regularization, obscures the
problem of time nonlocality. In general, however, the time nonlocality will
bring up some new terms which will be of importance for us here. We will see
this when we consider nucleon matrix elements of local quark operators.


\subsection{Nucleon matrix elements}

Let us consider a local quark operator $O$ defined in analogy with Eq.(\ref
{Ocoll}): 
\begin{equation}
{O}_\alpha ^\Gamma (t)=\overline{\psi }(t)\gamma _0\Gamma \lambda _\alpha
\psi (t)\;.
\end{equation}
We will calculate now the nucleon matrix element of the collective operator $%
{\cal O}_\alpha ^\Gamma $ in the path integral formalism: 
\begin{eqnarray}
\left\langle N,\vec{p}^{~\prime }\left| {\cal O}_\alpha ^\Gamma (t_1)\right|
N,\vec{p}\right\rangle  &=&\frac 1{{\cal Z}^{\prime }}\int d^3xd^3y\ e^{+i%
\vec{p}^{\;\prime }\cdot \vec{x}}e^{-i\vec{p}\cdot \vec{y}}\int {\cal D}\psi
^{\dagger }{\cal D}\psi {\cal D}U~~~~~~~~~~~~~~~~~  \nonumber \\
&&\ J_N(\vec{x},+T/2)\;{O}_\alpha ^\Gamma (t_1)\;e^{i\int d^4x^{\prime
}L}J_N^{\dagger }(\vec{y},-T/2)\ \ ,
\end{eqnarray}
where the ${\cal Z}^{\prime }$ is a normalization factor which gives: 
\[
\langle N,\vec{p}^{~\prime }|N,\vec{p}\rangle =(2\pi )^3\delta ^3(\vec{p}%
^{~\prime }-\vec{p})\ .
\]

After integrating over the quark fields and the collective coordinates
(rotational zero mode and also translational zero mode), we obtain 
\begin{eqnarray}
\left\langle N,\vec{p}^{~\prime }\left| {\cal O}_\alpha ^\Gamma (t_1)\right|
N,\vec{p}\right\rangle  &=&\frac 1{{\cal Z}^{\prime \prime }}\int d^3x\
e^{-i(\vec{p}^{\;\prime }-\vec{p})\cdot \vec{x}}\int {\cal D}\xi \ \Psi
_N^{\dagger }(T/2)\;  \nonumber \\
\left\{ D_{\alpha \beta }(t_1)\Lambda _\Gamma ^\beta (\vec{x})\right. 
&-&\int dt\left[ D_{\alpha \beta }(t_1)\;I_\Gamma ^{\beta \gamma }(\vec{x}%
,t_1-t_2)\;\Omega _\gamma (t_2)\ \Theta (t_1-t_2)\right.   \label{eq-timeord}
\\
&+&\left. \left. \Omega _\gamma (t_2)\;{I_\Gamma ^{\beta \gamma \;\dagger }}(%
\vec{x},t_1-t_2)\;D_{\alpha \beta }(t_1)\ \Theta (t_2-t_1)\right] \right\}
e^{i\int dt{\cal L}(t)}\;\Psi _N(-T/2)\ .  \nonumber
\end{eqnarray}
Note that time dependence of $D$ matrices, $\Omega $'s and wave functions $%
\Psi _N$ appears through time dependence of the Euler angles $\xi =\xi (t)$.
The {\it densities} introduced in Eq.(\ref{eq-timeord}) are defined by: 
\begin{eqnarray}
\Lambda _\Gamma ^\beta (\vec{x}) &=&N_{{\rm c}}\sum\limits_m\langle m|\vec{x}%
\rangle \Gamma \lambda _\beta \langle \vec{x}|m\rangle \ ,  \nonumber \\
I_\Gamma ^{\beta \gamma }(\vec{x},t) &=&i\frac{N_{{\rm c}}}4%
\sum\limits_{mn}e^{-i(\varepsilon _n-\varepsilon _m)t}\langle m|\vec{x}%
\rangle \Gamma \lambda _\beta \langle \vec{x}|n\rangle \langle n|\lambda
_\gamma |m\rangle \ ,  \nonumber \\
I_\Gamma ^{\beta \gamma \;\dagger }(\vec{x},t) &=&i\frac{N_{{\rm c}}}4%
\sum\limits_{mn}e^{i(\varepsilon _n-\varepsilon _m)t}\langle m|\lambda
_\gamma |n\rangle \langle n|\vec{x}\rangle \Gamma \lambda _\beta \langle 
\vec{x}|m\rangle \ .  \label{dens}
\end{eqnarray}

\begin{figure}[h]
\centerline{
\epsfig{file=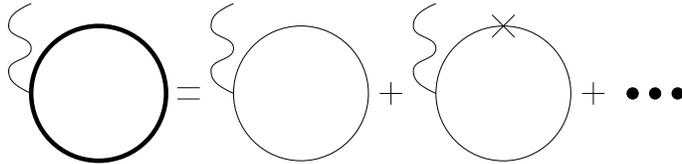,width=9cm}
}
\caption{Matrix element of the external current denoted by
a curly line expanded in powers of angular velocity
$\Omega$ (denoted by crosses).}
\label{Fig3}
\end{figure}

Equation (\ref{eq-timeord}) defines clearly the order of the operators $%
\Omega_\gamma$ and $D_{\alpha \beta}$. 
This is pictorially illustrated in Fig.\ref{Fig3}.
Only now, following the trick of
adiabatic approximation employed already in Eqs.(\ref{adiabatic},\ref{Ialbe}%
), and using (\ref{DOm}) we arrive at the formula for the matrix element of
the operator ${\cal O}^\Gamma_\alpha$ which has been already anticipated in
Eq.(\ref{Ocoll1}). Let us remark that the formulae obtained with the help of
Eqs.(\ref{dens}) are slightly more general -- they allow for calculation of
momentum dependent matrix elements, which we will need to calculate
form factors, for example. Integrated over $dx$ with $\vec{p}^{~\prime}=\vec{p%
}$ they reduce to (\ref{Ocoll1}).

Let us now apply our results to the operators of physical interest. In
section \ref{toy} we have already calculated the charge operator. Next, let
us consider axial vector current. To this end one considers an operator: 
\begin{equation}
{O}_{AK}^{{\rm ax.}}=\overline{\psi }\lambda _A\gamma _K\gamma _5\psi \;.
\end{equation}
Finally, we will consider magnetic moment operator:  
\begin{equation}
{\cal O}_{QK}^{{\rm mag.}}=\overline{\psi }\lambda _Q(\vec{r}\times \vec{%
\gamma})_K\psi \;.  \label{Omag}
\end{equation}
Here $Q$ corresponds to the electric charge: 
\begin{equation}
Q=\frac 12\lambda _3+\frac 1{2\sqrt{3}}\lambda _8\;.
\end{equation}
Note that operator (\ref{Omag}) depends on space variable $\vec{r}$ and it
involves nontrivial integration over $d^3x$.

Let us analyze various contributions entering Eq.(\ref{Ocoll1}) and define
the pertinent inertia parameters. From general symmetry considerations one
gets: 
\begin{eqnarray}
{\cal O}_{\alpha K}^\Gamma &=& D_{\alpha K} X_0  \nonumber \\
&- & \left\{ D_{\alpha \rho}, \Omega_\beta \right\} \; \left( \frac{1}{\sqrt{%
3}} \delta_{K \beta} \delta_{\rho 8} X_{1}^{(+)} + d_{K r b} X_{2}^{(+)}
\right)  \nonumber \\
&- & i \left[ D_{\alpha \rho}, \, \Omega_\beta \right] \; \left(-\frac{1}{2}
\epsilon_{K R B} X_{1}^{(-)} + f_{K r b} \; X_2^{(-)} \right) \ .
\label{Xdef}
\end{eqnarray}
Equation ({\ref{Xdef}) defines inertia parameters $X_{1,2}^{(\pm)}$ which
depend explicitly on the Lorentz structure $\Gamma$. They can be read off
directly from Eq.(\ref{Ocoll1}). After some algebra we get: 
\begin{eqnarray}
{\cal O}_{\alpha K}^\Gamma = \left( X_0 - \frac{X_1^{(-)}}{I_1^{(+)}} + 
\frac{X_2^{(-)}}{I_2^{(+)}} \right) D_{\alpha K}\;-\; \frac{2}{\sqrt{3}}\;
D_{\alpha 8} {\cal S}_K \; \frac{X_1^{(+)}}{I_1^{(+)}} \;-\;2
d_{pqK}D_{\alpha p} {\cal S}_q \; \frac{X_2^{(+)}}{I_2^{(+)}} \; .
\label{genform}
\end{eqnarray}
This form of ${\cal O}_{\alpha K}^\Gamma$ can be also used for the singlet
operators, {\em i.e.} for $\alpha=0$ provided one makes a replacement: 
\begin{equation}
D_{08} \rightarrow \sqrt{3} ~~~~~{\rm otherwise}~~~~~~ D_{0\beta}
\rightarrow 0 \; .
\end{equation}
}

Finally, let us write explicit forms of the inertia parameters both for the
axial current \cite{BPGaxial}: 
\begin{eqnarray}
A_{0} &=& N_{{\rm c}} \sum_m \langle m| \sigma_3 {\tau}_3 |m \rangle \ , 
\nonumber \\
A_{1}^{(+)} &=& \frac{N_c}{2} \sum_{m,n} \frac{1}{\varepsilon_n-\varepsilon_m%
} \langle m| \sigma_3 |n \rangle \langle n|\tau_3|m \rangle \ ,  \nonumber \\
A_{1}^{(-)} &=& i \frac{N_c}{2} \sum_{m,n} \frac{1}{\varepsilon_n-%
\varepsilon_m} \langle m| (\vec{\sigma}\times\vec{\tau})_3 |n \rangle
\langle n|\tau_3|m \rangle \ ,  \nonumber \\
A_{2}^{(\pm)} &=& \frac{N_c}{6} \left\{ \sum_{m,n_0} \frac{\langle m| \vec{%
\sigma}\cdot\vec{\tau} |n_0 \rangle \langle n_0|m \rangle}{%
\varepsilon_{n_0}-\varepsilon_m} \pm \sum_{m_0,n} \frac{\langle m_0| \vec{%
\sigma}\cdot\vec{\tau} |n \rangle \langle n|m_0 \rangle}{\varepsilon_n-%
\varepsilon_{m_0}} \right\}  \label{Aiall}
\end{eqnarray}
and for magnetic moments \cite{KPBGmag}: 
\begin{eqnarray}
M_{0} &=& \frac{N_{{\rm c}}}{6} \sum_m \langle m|\gamma_5 (\vec{r} \times 
\vec{\sigma})\cdot\vec{\tau} |m \rangle \ ,  \nonumber \\
M_{1}^{(+)} &=& \frac{N_{{\rm c}}}{4} \sum_{m,n} \frac{1}{%
\varepsilon_n-\varepsilon_m} \langle m|\gamma_5 (\vec{r} \times \vec{\sigma}%
)_3 |n \rangle \langle n|\tau_3|m \rangle \ ,  \nonumber \\
M_{1}^{(-)} &=& i \frac{N_{{\rm c}}}{4} \sum_{m,n} \frac{1}{%
\varepsilon_n-\varepsilon_m} \langle m| ((\vec{r}\times\vec{\sigma})\times 
\vec{\tau})_3 |n \rangle \langle n|\tau_3|m \rangle \ ,  \nonumber \\
M_{2}^{(\pm)} &=& \frac{N_{{\rm c}}}{12} \left\{ \sum_{m,n_0} \frac{\langle
m| \gamma_5 (\vec{r} \times \vec{\sigma})\cdot\vec{\tau} |n_0 \rangle
\langle n_0|m \rangle}{\varepsilon_{n_0}-\varepsilon_m} \pm \sum_{m_0,n} 
\frac{\langle m_0|\gamma_5 (\vec{r} \times \vec{\sigma})\cdot\vec{\tau}|n
\rangle \langle n|m_0 \rangle}{\varepsilon_n-\varepsilon_{m_0}} \right\} \ .
\label{Miall}
\end{eqnarray}

Our analysis in this section has been based on the unregularized effective
action (\ref{ourS}). For practical calculations one has to employ
technically more involved methods in order to find the explicit form of the
regulators. The relevant formulae can be found in the literature 
\cite{BPGaxial,KPBGmag}. 
However, not all inertia parameters require regularization. In
the case of the axial current and the magnetic moment operator $A_{1,2}^{(+)}
$ and $M_{1,2}^{(+)}$ do not have to be regularized. They follow from the
imaginary part of the action in the Euclidean formulation of the model and
they have counterparts in the Skyrme model \cite{WaSkyrme}
where they are related to the
Wess-Zumino term. Quantities with superscript  ''$(-)$'' do not have
counterparts in the Skyrme model\footnote{%
Precisely speaking in the simplest version of the Skyrme model without
vector mesons.}. They are related to the non-locality of the effective action.


\section{Small soliton limit}

\label{smallsol}

We shall discuss now the collective operators for the axial couplings and
magnetic moments. Collective, in the sense that they have to be sandwiched
between the baryon wave functions (\ref{wf}). Equation (\ref{genform})
together with the expression for the charge (given already in Eqs.(\ref{TS2},%
\ref{eq-ecbk})) suffer from the quantization dilemma discussed in the
introduction. In this section we will study the behavior of various terms
in Eqs.(\ref{genform}) and (\ref{eq-ecbk}) in the limit of the small soliton.

Chiral Quark-Soliton Model interpolates between the naive nonrelativistic
quark model and the Skyrme model. To this end one studies the model with
fixed parameters (constituent quark mass $M$ and cutoff parameters) by
varying the soliton profile function $F(r)$. If the soliton size $r_0$
(defined through $F=F(r/r_0)$) decreases, the valence level joins the upper
continuum and the {\it sea} contribution vanishes. For large solitons the
valence level sinks into the Dirac sea and the explicit contribution of the
valence level disappears. In Ref.\cite{smallsol} it was shown that in the
SU(2) version of the model the well known quark model result $g_{{\rm A}%
}^{(3)}=(N_{{\rm c}}+2)/3$ can be recovered only if the rotational
correction proportional to $A_1^{(-)}/I_1^{(+)}$ is taken into account. This
is a convincing argument in favor of the rotational corrections which
originate from the time nonlocality of the quark loop. Moreover, in the
limit of large soliton size this correction is dying out much faster than
the leading contribution. This result is also welcome, since in this limit $%
\chi$QSM results should reduce to the ones of the Skyrme model. Indeed, in
the Skyrme model this kind of rotational corrections does not exist, because
the model is based on a local Lagrangian density.

Here we are going to reverse the logic used in Ref.\cite{smallsol}. We will
investigate the small soliton size of the SU(3) formulae (\ref{genform}) in
order to identify terms which spoil the agreement with the quark model
results. We will also demand that both versions of the model SU(2) and SU(3)
have the same limit for small solitons. These two requirements suffice to
identify unambiguously the troublesome terms. As we shall see these terms
correspond to the quantities $X_2^{(-)}$ of Eq.(\ref{Xdef}).

The nonrelativistic quark model (NQM) limit is most naturally defined in
terms of a somewhat unphysical profile function $F=\pi \theta (r_0-r)$ with $%
r_0\rightarrow 0$. In this limit the energy of the valence level $E_{\val%
}\rightarrow M$ and the {\it sea} contributions to all quantities vanish.
Simultaneously we take the nonrelativistic limit, i.e. we systematically
neglect small components of the spinors $|\;i\;\rangle $. Although the
inertia parameters $X_{1,2}^{(\pm )}$ are infinite in this limit because of
the energy denominators, their ratios to the moments of inertia $%
I_{1,2}^{(+)}$ remain finite. In particular: 
\begin{equation}
\frac{I_2^{(-)}}{I_2^{(+)}}\rightarrow 1
\end{equation}
and the nucleon charge (\ref{eq-ecbk}) has wrong NQM limit as $%
r_0\rightarrow 0$.

Let us next consider the axial coupling $g_{{\rm A}}^{(3)}$. To this end we
have to calculate matrix elements of the collective operators involving
Wigner $D$ functions in the state of a proton with spin up. The results for
SU(2) and SU(3) take the following form: 
\begin{eqnarray}
g_{{\rm A}}^{(3)}[{\rm SU(2)}] &= & -\frac{1}{3} \left( A_0 - \frac{A_1^{(-)}%
}{I_1^{(+)}} \right) ,  \nonumber \\
g_{{\rm A}}^{(3)}[{\rm SU(3)}] &= & -\frac{7}{30} \left( A_0 - \frac{%
A_1^{(-)}}{I_1^{(+)}} + \frac{A_2^{(-)}}{I_2^{(+)}} \right) \;-\; \frac{1}{30%
}\frac{A_1^{(+)}}{I_1^{(+)}} \;-\; \frac{7}{30} \frac{A_2^{(+)}}{I_2^{(+)}}
\; .  \label{gasu23}
\end{eqnarray}

Next we consider NQM limit of the quantities entering Eq.(\ref{gasu23}): 
\begin{eqnarray}
A_0 \rightarrow -3 \ , ~~~~ \frac{A_{1}^{(+)}}{I_1^{(+)}} \rightarrow -1 \ ,
~~~~ \frac{A_{1}^{(-)}}{I_1^{(+)}} \rightarrow 2 \ , ~~~~ \frac{A_{2}^{(\pm)}%
}{I_2^{(+)}} \rightarrow -2 \; .  \label{Alimit}
\end{eqnarray}
With these values we obtain: 
\begin{eqnarray}
g_{{\rm A}}^{(3)}[{\rm SU(2)}]\rightarrow \frac{5}{3} &~~~~~~{\rm and}%
~~~~~~~ & g_{{\rm A}}^{(3)}[{\rm SU(3)}]\rightarrow \frac{5}{3}+\frac{7}{15}
\ ,  \label{galimit}
\end{eqnarray}
where $7/15$ for $g_{{\rm A}}^{(3)}$[SU(3)] is the limiting value of ${%
A_{2}^{(-)}}/{I_2^{(+)}}$.

Finally we will consider magnetic moments of proton and neutron: 
\begin{eqnarray}
\mu _{{\rm N}}[{\rm SU(2)}] &=&-\frac 16\,\frac{M_1^{(+)}}{I_1^{(+)}}-\frac 1%
6\left[ M_0-\frac{M_1^{(-)}}{I_1^{(+)}}\right] 2\,T_3\;,  \nonumber \\
\mu _{{\rm N}}[{\rm SU(3)}] &=&-\frac 1{60}\left[ \left( M_0-\frac{M_1^{(-)}%
}{I_1^{(+)}}+\frac{M_2^{(+)}+M_2^{(-)}}{I_2^{(+)}}\right) +3\frac{M_1^{(+)}}{%
I_1^{(+)}}\right]   \nonumber \\
&&-\frac 1{60}\left[ 7\left( M_0-\frac{M_1^{(-)}}{I_1^{(+)}}+\frac{%
M_2^{(+)}+M_2^{(-)}}{I_2^{(+)}}\right) +\frac{M_1^{(+)}}{I_1^{(+)}}\right]
2\,T_3\;.  \label{muNsu23}
\end{eqnarray}
Inertia parameters are given in terms of the space integral $K$ which can be
found in the Appendix. In the limit of the small soliton size we get: 
\[
M_0\rightarrow -2K\ ,~~~~\frac{M_1^{(+)}}{I_1^{(+)}}\rightarrow -\frac 23K\
,~~~~\frac{M_1^{(-)}}{I_1^{(+)}}\rightarrow \frac 43K\ ,~~~~\frac{M_2^{(\pm
)}}{I_2^{(+)}}\rightarrow -\frac 43K\;.
\]
With these values we obtain: 
\begin{equation}
\mu _N[{\rm SU(2)}]\rightarrow \frac 19(1\pm 5)K~~~~~~{\rm and}~~~~~~~\mu _N[%
{\rm SU(3)}]\rightarrow \frac 19(1\pm 5)K+\frac 1{45}(1\pm 7)K\;,
\label{maglim}
\end{equation}
where $+$ refers to the proton and $-$ to the neutron, respectively. The
last term in Eq.(\ref{maglim}) corresponds to the limiting value of $%
M_2^{(-)}/I_2^{(+)}$. It can be readily seen from Eq.(\ref{maglim}), that
the quark model ratio ${\mu _{{\rm p}}}/{\mu _{{\rm n}}}=-3/2$ is recovered
only if this term is neglected. Indeed: 
\begin{equation}
\frac{\mu _{{\rm p}}}{\mu _{{\rm n}}}\rightarrow -\frac 32+\frac 1{13},
\end{equation}
where $1/13$ corresponds to $M_2^{(-)}/I_2^{(+)}$.

In this section we have calculated small soliton limit of the collective
operators corresponding to charge, axial decay constants and magnetic
moments. In the SU(2) case the limiting values coincide with the ones of
the nonrelativistic quark model. In the case of SU(3) they get extra
contributions coming from $X_2^{(-)}$ defined in Eq.({\ref{Xdef}). If we
neglect these terms the nonrelativistic quark model values are recovered.
Let us stress that the agreement with the quark model results {\em cannot}
be obtained without the contributions of $X_1^{(-)}$, which originate from
the time non-locality of the fermion loop as $X_2^{(-)}$. There reason why 
$X_2^{(-)}$ do not vanish, is the isospin embedding of the SU(2) {\em %
hedgehog} in the SU(3) $U$ field given by Eq.(\ref{U3}). Because of this
asymmetric embedding the Dirac spectrum (\ref{spectrum}) is asymmetric, and,
as a result, $X_2^{(-)}$ are non-zero. In the case of the charge operator it
is relatively easy to argue, as it was done in the introduction, that $%
I_2^{(-)}/I_2^{(+)}$ corresponds to the nonleading $1/N_{{\rm c}}$
correction to the coefficient in front of ${\cal S}_8$. However, it is not
possible to make such a clear distinction for other quantities.
Nevertheless, in the following we will neglect $X_2^{(-)}$ terms. In that
way the agreement with the nonrelativistic quark model will be restored and,
as we will see in the next section, we will be able to describe the
experimental data with much better degree of accuracy. }

Let us finally note that even if we discard terms proportional to $X_2^{(-)}$
the SU(3) expressions, in comparison to SU(2), contain extra terms, as seen
explicitly from Eqs.(\ref{gasu23}) and (\ref{muNsu23}). Nevertheless for $%
r_0 \rightarrow 0 $ the limiting values for both SU(2) and SU(3) do
coincide. This is due to the fact that the matrix elements of the $D$
functions are much smaller in the case of SU(3) and the extra terms make up
for this difference, so that finally the SU(2) and SU(3) limiting values
are the same. This will be also almost exactly true for the realistic
self-consistent soliton profile.

\section{Numerical results}

\label{numres}
 
The method we use for the numerical computation is based on the sums over 
quark single-particle levels in the background pion field.
The system is in a spherical 3-dimensional box where the eigenfunctions of 
the Dirac Hamiltonian in the background pion field can be obtained by 
numerical diagonalization~\cite{KahaRip}.
For the ultraviolet regularization of the Dirac sea contribution we employ the 
proper-time scheme. The ultraviolet cutoff $\Lambda$ is determined by 
fitting the pion decay constant and it reads: $\Lambda\sim 600$MeV.
The soliton profile is found by self-consistent minimization of the static 
classical energy. All results are for $m_{\pi}=m_{\rm K}=140$~MeV.

\begin{figure}[htb]
\epsfig{file=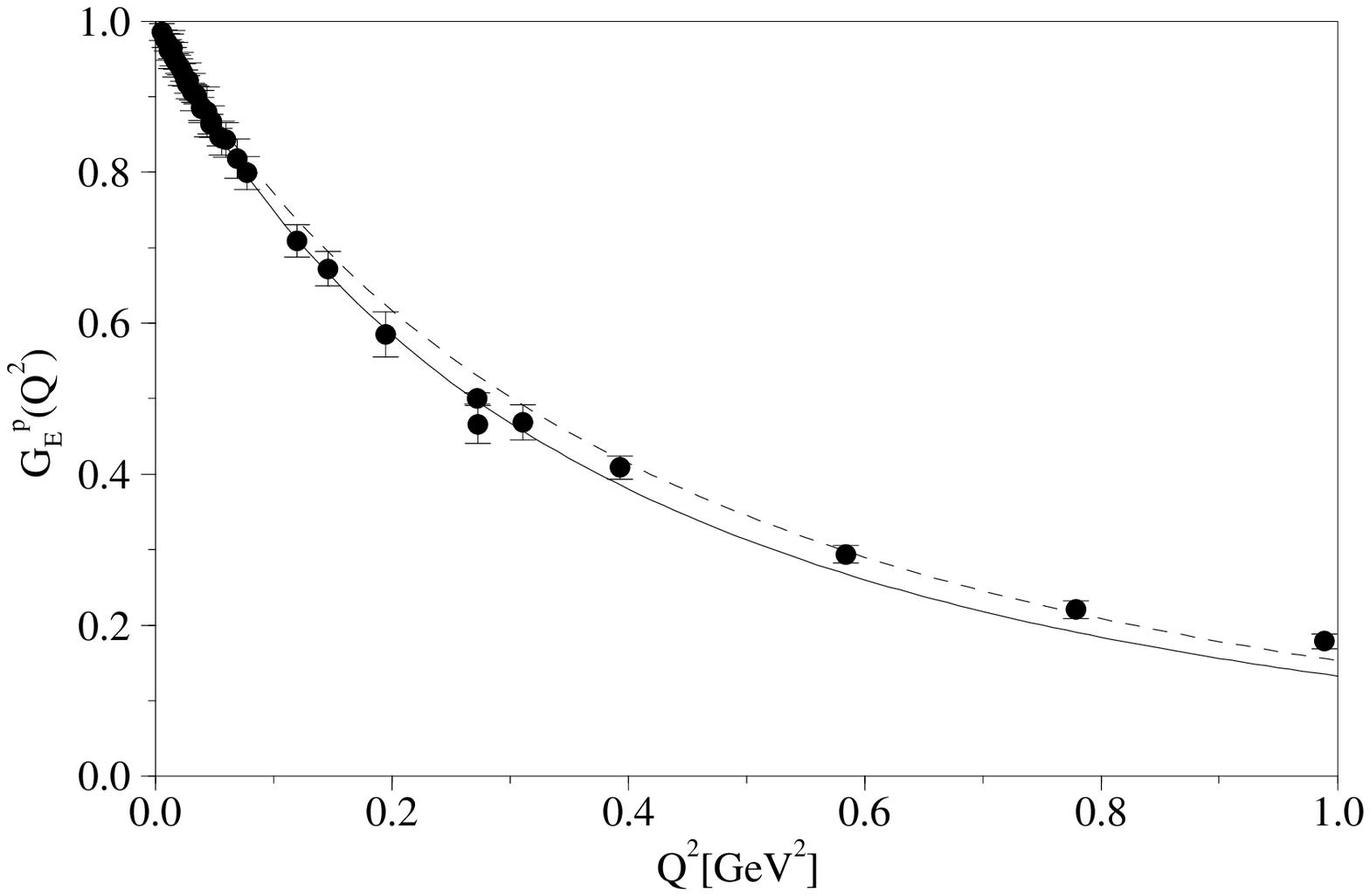,width=10cm}
\caption{Proton electric form factor in the SU(2) (dashed line) and 
sym. cons. SU(3) (solid line) version of the model. 
The experimental data are from ref.~{\protect\cite{Hoh}}.}
\label{fig:elp}
\end{figure}

\begin{figure}[htb]
\epsfig{file=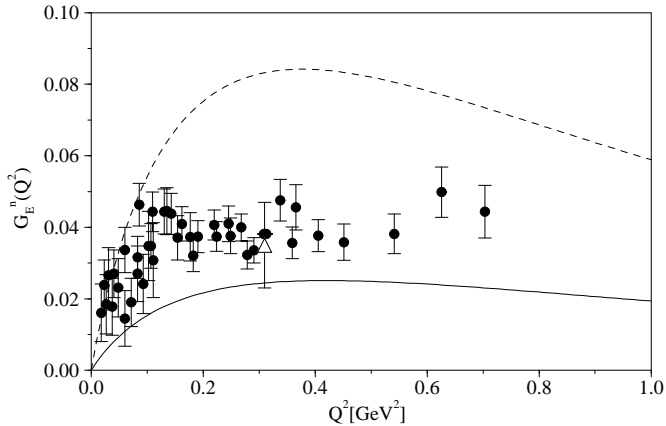,width=10cm}
\caption{Neutron electric form factor in the SU(2) (dashed line) and 
sym. cons. SU(3) (solid line) version of the model.
The experimental data are from ref.~{\protect\cite{Pla}} denoted by 
solid circles and ref.~{\protect\cite{Mey}} denoted by an open triangle.}
\label{fig:eln}
\end{figure}

\begin{figure}[htb]
\epsfig{file=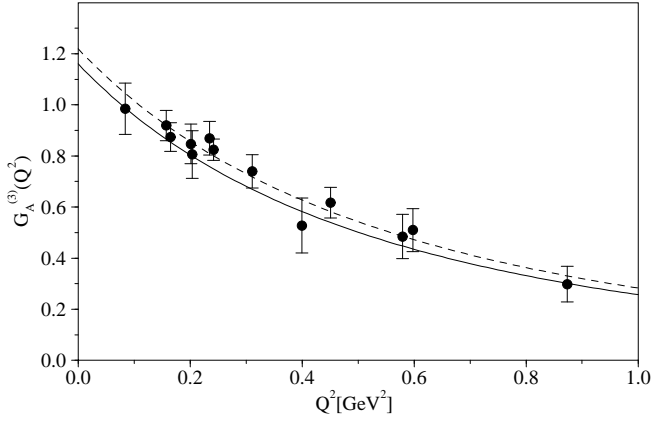,width=10cm}
\caption{Nucleon axial form factor in the SU(2) (dashed line) and 
sym. cons. SU(3) (solid line) version of the model.
The experimental data are from ref.~{\protect\cite{Bak,Kit}}.}
\label{fig:ax}
\end{figure}

\begin{figure}[htb]
\epsfig{file=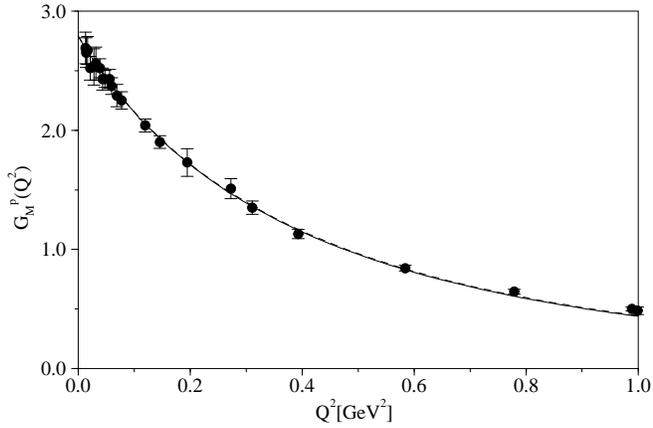,width=10cm}
\caption{Proton magnetic form factor in the SU(2) (dashed line) and 
sym. cons. SU(3) (solid line) version of the model.
The experimental data are from ref.~{\protect\cite{Hoh}}.}
\label{fig:mgp}
\end{figure}

\begin{figure}[htb]
\epsfig{file=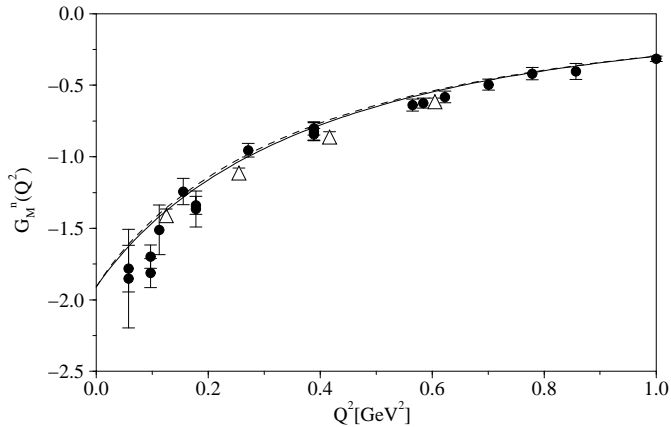,width=10cm}
\caption{Neutron  magnetic form factor in the SU(2) (dashed line) and 
sym. cons. SU(3) (solid line) version of the model.
The experimental data are from ref.~{\protect\cite{Hoh}} denoted by 
solid circles and ref.~{\protect\cite{Bru}} denoted by open triangles.}
\label{fig:mgn}
\end{figure}

Before considering SU(3) results within the prescription which we have 
developed in this work it is instructive to review the SU(2) results given 
in Ref.~\cite{BOrev}
Fig.~\ref{fig:elp} shows the proton electric form factor calculated with 
$M=420$~MeV.
The dashed curve corresponds to the SU(2) result and it follows
the experimental data with very good accuracy.
In Fig.~\ref{fig:eln} we plot neutron electric form factor.
The dashed curve corresponding to the SU(2) version of the model 
overestimates the data. We will come back to this point later.
In Fig.~\ref{fig:ax} we plot the axial form factor of the nucleon.
It is also in a good agreement with the experiments.
In this case the next-to-leading order correction of
 $1/N_c$ expansion of the rotational zero-mode of soliton,
namely $A_1^{(-)}/I_1^{(+)}$ in Eq.(\ref{gasu23}), is 
crucial to reproduce the experimental data~\cite{WaWa,allstars}.
Moreover this correction ensures that in the limit of the small soliton
our result agrees with the estimation of the naive non-relativistic 
quark model: $g_{\rm A }^{(3)}=(N_{\rm c}+2)/{3}$ ~\cite{smallsol}.
In Figs.~\ref{fig:mgp} and \ref{fig:mgn} the magnetic form factors of proton 
and neutron are shown.
For the magnetic properties we also take into account the next-to-leading order
correction of $1/N_{\rm c}$ expansion, namely $M_1^{(-)}/I_1^{(+)}$ in 
Eq.(\ref{muNsu23}).
The theoretical curves are multiplied with a factor such that the values at 
$Q^2=0$ agree with the corresponding experimental magnetic moments.
Apparently the $q$-dependence of the form factors is well reproduced.
The theoretical magnetic moments deviate noticeably from the experimental ones,
as one can see at Table.
This unfortunate feature is in accordance with most, if not all chiral models 
which notoriously have difficulties to obtain proper absolute values of the 
magnetic moments.

\begin{table}
\begin{tabular}{cccc}
              & SU(2) & sym. cons. SU(3) & exp. \\
$\mu_p$[n.m.] &  1.98 &  1.81           &  2.79 \\
$\mu_n$[n.m.] & -1.36 & -1.20           & -1.91
\end{tabular}
\end{table}

The solid curves in Figs.~\ref{fig:elp}-\ref{fig:mgn} correspond to the SU(3) 
results truncated according to the prescription advocated in 
section \ref{smallsol}.
The differences between the results in SU(2) and SU(3) are very small, 
except the case of the neutron electric form factor, Fig.~\ref{fig:eln}.

In the $\chi$QSM  all quantities consists from two kinds of contributions.
One is the valence level contribution, and the other one  corresponds to the 
Dirac sea levels. The proton electric form factor is dominated by the 
valence level contribution, while the neutron one is dominated  by the Dirac 
sea, which is interpreted as contribution of a mesonic cloud.
The remarkable difference between SU(2)- and SU(3) calculation exists only for 
the neutron electric form factor and there are two possible explanations:
1) It might be due to a systematic error of the model.
Because the neutron electric form factor is quite tiny compared with the one 
of the proton, 
it is quite probable  that it is sensitive to the  systematic error of the model.
2) It might be due to a kaonic excitations of the vacuum.
In the SU(2) the vacuum excitation (Dirac sea polarization) in the neutron 
occurs only in the $\pi^-$ channel.
On the other hand, in the SU(3), it has another  contribution corresponding 
to the $K^+$ channel. The $K^+$ excitation suppresses the $\pi^-$ excitation.
Because of that, the SU(3) result is below the result in SU(2). One should also 
stress the sensitivity of the magnetic form factors to the tail of the
soliton field. 

In this paper we have treated strange degrees of freedom in a purely 
perturbative way. One possible method which assumes strong SU(3) breaking 
is given by the bound-state approach proposed by Callan and
Klebanov in the Skyrme model~\cite{CK}. We plan to investigate the neutron 
electric form factor with a method based on 
the bound-state approach to the kaon.

\section{Summary}

\label{disc}

The aim of this paper was to discuss in full extent the quantization paradox,
which arises, when one applies canonical quantization rules to the rotating
soliton in the SU(3) Chiral Quark-Soliton Model. We have argued, that
canonical quantization leads to the two type of terms, called 
$X_{1,2}^{(-)}$, which appear due to the non-commutativity of the 
SU(3) Wigner functions $D_{\alpha\rho}$ and the angular
velocities $\Omega_\beta$, and the non-locality of the fermion loop. These
terms constitute $1/N_{\rm c}$ corrections to the physical observables in
question, and, at first sight, it seems that in the consistent treatment either
both types should be simultaneously retained or neglected. However, as
we have shown in full detail,  quantities $X_{1}^{(-)}$, which involve only 
the transitions between the states from the SU(2) soliton subspace are 
theoretically harmless and  phenomenologically desperately needed, whereas 
terms $X_{2}^{(-)}$, which involve the transitions between 
the soliton states and plane waves lead to the violation of the canonical
quantization rules for the flavor generators. So far
no satisfactory theoretical solution of this paradox has been found. 
We have proposed a semi-phenomenological way, based on the requirement
that the $\chi$QSM model results should agree with the non-relativistic
quark model in the limit of the artificially small soliton, to circumvent the
apparent contradiction and discard the $X_{2}^{(-)}$ terms. The only
justification of this procedure, which we can give at the moment, is based
on the fact, that the SU(3) canonical quantization mixes different orders
of $1/N_{\rm c}$ expansion and it is a priori not clear which terms are
only formally subleading, and which constitute genuine corrections to
the physical quantities. We have discussed this at length in the Introduction 
on the example of the nucleon charge, where such a distinction is easy
to make. Unfortunately other observables like axial couplings and magnetic
moments, which we have discussed in this paper, are much more complicated,
and the same kind of argument cannot be straightforwardly applied in that
case.

Once one accepts the philosophy that  $X_{1}^{(-)}$ terms should be kept
as they come out from the calculations and $X_{2}^{(-)}$ terms should be
discarded, a consistent picture of the SU(3) soliton emerges. All collective
operators obey proper commutation rules, and all observables have proper
(i.e. non-relativistic quark model) limit for very small solitons. The SU(2)
and SU(3) results coincide and the agreement with the experiment is,
generally speaking, satisfactory.

We felt obliged to discuss this problem without being able to give sound
theoretical solution to it, since it plagues SU(3) model calculations published
in the literature. We feel that non-expert readers should be aware of 
this problem, with the hope that some of them may find an elegant explanation
in a not distant future.

\section*{Acknowledgments}

We thank W. Broniowski and  G. Ripka for discussions and many remarks
which we have included in this paper. Special thanks are due to P.V.
Pobylitsa and M.V. Polyakov for their interest and criticism.
This work has been partially supported by Polish-German collaboration
agreement between Deutsche Forschungsgemeinschaft and Polish Academy
of Sciences. M.P. acknowledges support of A. v. Humboldt Foundation.


\end{document}